\documentclass[aps,prd,11pt,superscriptaddress,notitlepage,longbibliography,nofootinbib,tightenlines]{revtex4-1}

\usepackage{amsmath,amssymb,amsfonts}
\usepackage{graphicx}
\usepackage{subfigure}
\usepackage{color}
\usepackage{hyperref}
\definecolor{darkred}{rgb}{0.8,0.1,0.1}
\hypersetup{colorlinks=true, linkcolor=darkred, citecolor=blue, }

\newcommand{\dS}{\texorpdfstring{\lowercase{d}S}{dS}}
\newcommand{\ii}{\ensuremath{\dot\imath}}
\newcommand{\scri}{\ensuremath{\mathcal{I}}}

\newcommand{\SO}[1]{\ensuremath{\mathsf{SO}(#1)}}
\newcommand{\SU}[1]{\ensuremath{\mathsf{SU}(#1)}}
\renewcommand{\O}[1]{\ensuremath{\mathsf{O}(#1)}}
\newcommand{\Sp}[1]{\ensuremath{\mathsf{Sp}(#1)}}

\DeclareMathOperator{\sech}{sech}
\DeclareMathOperator{\diag}{diag}
\DeclareMathOperator{\partialLR}{\tensor{\partial}}

\makeatletter\def\l@subsubsection#1#2{}%
\makeatother

\begin{document}

\title{Higher-spin realization of a dS static patch/cut-off CFT correspondence}

\author{Andreas Karch}
\email{akarch@u.washington.edu}
\affiliation{Department of Physics, University of Washington, Seattle, WA 98195-1560, USA}

\author{Christoph F.~Uhlemann}
\email{uhlemann@physik.uni-wuerzburg.de}
\affiliation{Institut f\"ur Theoretische Physik und Astrophysik, Universit\"at W\"urzburg, 97074 W\"urzburg, Germany}

\begin{abstract}
We derive a holographic relation for the dS static patch with the dual field theory defined on the observer horizon.
The starting point is the duality of higher-spin theory on AdS$_4$ and the \O{N} vector model.
We build on a similar analytic continuation as used recently to obtain a realization of dS/CFT,
and adapt it to the static patch.
The resulting duality relates higher-spin theory on the dS$_4$ static patch to 
a cut-off CFT on the cylinder $\mathbb{R}\,{\times}\,$S$^2$.
The construction permits a derivation of the finite thermodynamic entropy associated to the 
horizon of the static patch from the dual field theory.
As a further brick we recover the spectrum of quasinormal frequencies 
from the correlation functions of the boundary theory.
In the last part we incorporate the dS/dS correspondence as an independent proposal for 
holography on dS and show that a concrete realization can be obtained by similar reasoning.
\end{abstract}


\maketitle
\tableofcontents

\section{Introduction}
The AdS/CFT correspondence \cite{Maldacena:1997re} has stimulated remarkable progress in the understanding of gauge theories.
Moreover, it provides a means to study quantum aspects of gravity with asymptotically-anti de-Sitter (AdS) boundary conditions
in terms of the dual conformal field theory (CFT).
However, our universe is likely not asymptotically AdS.
It would therefore be desirable to have a holographic definition of (quantum) gravity 
in terms of a dual boundary theory also on the physically more directly relevant de Sitter (dS) space.
There has indeed been a proposal for a dS/CFT correspondence \cite{Strominger:2001pn},  which exploits the 
conformal properties of dS to establish a dual CFT description on the spacelike
conformal boundary at future/past infinity $\scri^\pm$. 
With the explicit realization obtained in \cite{Anninos:2011ui} this proposal has recently been lifted to a very concrete level.
However, with the dual CFT defined at $\scri^\pm$ these dS/CFT correspondences are formulated in terms
of dS meta observables, accessible only to an unphysical meta observer \cite{Witten:2001kn}.

Restricting to only the region accessible to a physical observer crucially complicates things.
There is no notion of a conformal boundary and we instead only have the horizon and the 
observer worldline as distinguished places. 
Moreover, this region is symmetric only under a subgroup of the dS isometries. 
Nevertheless, understanding quantum gravity on the region accessible to a single
observer arguably is the most interesting question to pose. 
We will therefore aim to explicitly realize holography in that setting.
Motivation for the existence of a holographic description comes in the first place from 
the Bekenstein bound, which certainly suggests that there is a holographic description also 
for gravity on the dS static patch.
The screen may in principle be anywhere. 
For the flat slicing of dS the conformal boundary is 
a preferred place since the full dS isometry group nicely acts on it.
With that option unavailable for the static patch the possibility of a dual quantum mechanics 
description on the observer worldline has been investigated in \cite{Anninos:2011af}.
However, the covariant construction of holographic screens in generic spacetimes \cite{Bousso:1999cb} 
suggests that the screen is at the horizon. 

In this note we aim to make this discussion more precise.
Similarly to \cite{Anninos:2011ui} we shall start from AdS/CFT dualities involving higher-spin 
theories \cite{Vasiliev:1990en,Vasiliev:1999ba} in the bulk, which may be seen as tensionless limits of string theory. 
More precisely, the bulk theory is the parity-invariant minimal bosonic version of Vasiliev gravity, with
massless symmetric tensor fields of all even spins.
We will exploit that there is a nice analytic continuation from AdS to dS for that theory \cite{Iazeolla:2007wt}.
This will allow us to derive from the well-understood Giombi-Klebanov-Polyakov-Yin AdS/CFT duality \cite{Klebanov:2002ja,Giombi:2009wh} 
by a double Wick rotation a dual description for higher-spin gravity on the static patch of dS$_4$.
The dual theory will be defined on the observer horizon
and will be a cut-off version of  the \Sp{N} CFT$_3$ of anticommuting scalars \cite{Robinson:2009xm},
which was obtained as dual theory at $\scri^+$ in \cite{Anninos:2011ui}. 
Without the geometric bells and whistles which are at the heart of the more conventional (A)dS/CFT dualities, 
establishing an analog of the bulk-boundary dictionary is a bit more subtle.
Building on the discussion of horizon holography in \cite{Sachs:2001qb} we will work out in detail how such
a dictionary can be realized for the static patch of dS and present some first applications.
We then turn to the dS/dS correspondence proposed in \cite{Alishahiha:2004md,Alishahiha:2005dj},
where the dual theories are similarly defined at a horizon.
It provides an independent approach to dS holography and we will adapt our construction to also obtain a concrete realization.

The outline is as follows. In Sec.~\ref{sec:dS-AdS-relation} we discuss an analytic continuation relating the dS static patch to an
inner shell of Euclidean AdS.
In Sec.~\ref{sec:staticpatch-holography} we build on the role of the AdS radial coordinate as an energy scale in the dual CFT 
to analytically continue a cut-off version of AdS/CFT to a holographic relation for the static patch.
This will first be restricted to an inner region before we recover the entire static patch in \ref{sec:holography-for-static-patch}.
The static-patch entropy will be derived from the dual theory in \ref{sec:entropy}. 
We then study the duality more explicitly for bulk scalar fields in Sec.~\ref{sec:scalars-explicitly}.
The realization of dS/dS correspondences will be derived in Sec.~\ref{sec:dsds}
and we conclude in Sec.~\ref{sec:discussion}.

\section{The {\dS} static patch as part of A\dS}\label{sec:dS-AdS-relation}

After reviewing the analytic continuation from dS to AdS as used in \cite{Anninos:2011ui}, 
we will in this section derive a similar relation of the dS static patch to an inner shell 
of AdS.
We start off from the well-known fact that dS$_{d+1}$ of signature $(-,+\dots +)$ and Euclidean AdS can both be defined
as hyperboloids in $(d\,{+}\,2)$-dimensional flat space with metric $\eta=\diag(-1,1,..,1)$ by
\begin{align}\label{eqn:hyperboloids}
 \text{dS}_{d+1}:\quad -X_0^2+\sum_{i=1}^{d+1}X_i^2=H^2~,&
 &\text{AdS}_{d+1}:\quad -X_0^2+\sum_{i=1}^{d+1}X_i^2=-L^2~.
\end{align}
Correspondingly, their symmetry groups \SO{1,d\,{+}\,1} coincide.
The defining equations are related by $H=\ii L$ and this can be 
exploited to relate their coordinatizations as follows.
The usual Poincar\'{e} coordinates can be introduced on AdS$_{d+1}$
by solving (\ref{eqn:hyperboloids}) in terms of 
\begin{align}\label{eqn:Poincare-coords}
 X_{0/1}&=\frac{u}{2}\left(1+\frac{1}{u^2}(\vec{x}^2\pm L^2)\right)~,&
 X_i&=\frac{Lx^{i-1}}{u}~,\quad \forall i=2,..,d+1~,
\end{align}
which results in the line element $ds^2=L^2 u^{-2}\big(du^2+d\vec{x}^2\big)$. 
The coordinates cover all of Euclidean AdS and $u=0$ corresponds to almost all of the conformal boundary.
The flat slicing of Lorentzian dS space, which covers half of the hyperboloid,
can now be obtained by the analytic continuation used in~\cite{Anninos:2011ui},
\begin{equation}\label{eqn:WickRotationPoincare}
 L=\ii H~,\qquad u=\ii \eta~.
\end{equation}
That results in $X_{0/1}\rightarrow \ii X_{0/1}$, such that their roles are exchanged and
we are dealing with a double Wick rotation in the ambient space. 
The resulting line element $ds^2=H^2 \eta^{-2}\big(-d\eta^2+d\vec{x}^2\big)$
corresponds to the flat slicing of dS$_{d+1}$ where $u$ has become the time coordinate.
Building on that simple geometric identification along with the analytic continuation from
AdS to dS for Vasiliev's higher-spin theory via (\ref{eqn:WickRotationPoincare}), a concrete 
realization of dS/CFT has been derived in \cite{Anninos:2011ui}.

\begin{figure}[ht]
\center
\subfigure[][]{ \label{fig:ds-ads1a}
  \includegraphics[width=0.39\linewidth]{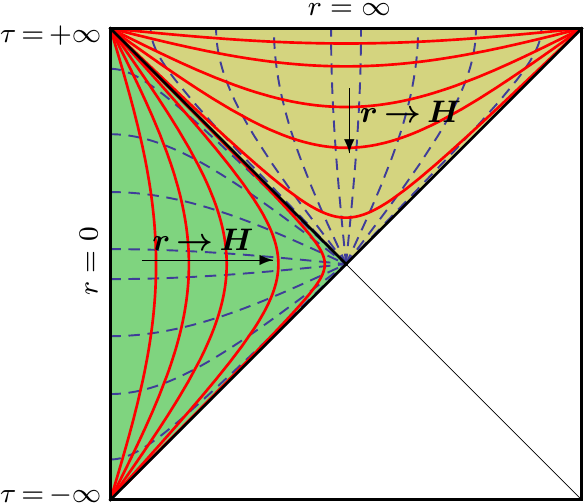}

}\qquad\qquad
\subfigure[][]{ \label{fig:ds-ads1b}
    \includegraphics[width=0.32\linewidth]{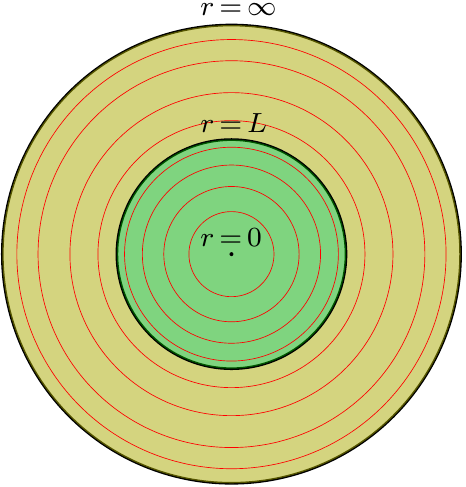}
}
\caption{The static patch of global dS and the extension to $r\,{>}\,H$ covering the expanding Poincar\'{e} patch
         are shown on the left hand side.
         The solid red and dashed blue curves correspond to constant $r$ and $\tau$, respectively.
	 Note the exchange of timelike and spacelike character for these curves when extending to $r\,{>}\,H$.
	 On the right hand side is a section of constant $\tau$, i.e.\ of constant latitude, 
	 through Euclidean AdS realized as the interior of a unit ball with coordinates (\ref{eqn:metric-EAdS-global}).
	 Thanks to the Killing vector field $\partial_\tau$ the sections are all equivalent.
	 The analytic continuation (\ref{eqn:WickrotationsGlobal}) identifies
         the inner green and outer yellow regions
         with the left and upper triangles of dS as shown on the left hand side, respectively.
         \label{fig:ds-ads1}
        }
\end{figure}

We will now discuss a similar relation of the dS static patch to an inner shell of
Euclidean AdS. To this end we turn to a different global coordinatization of Euclidean AdS.
This is obtained by solving (\ref{eqn:hyperboloids}) in terms of 
\begin{align}\label{eqn:EAdS-global}
 X_0&=\sqrt{L^2+r^2}\cosh\tau~,&
 X_1&=\sqrt{L^2+r^2}\sinh\tau~,&
 X_i=r z_i\quad\forall i=2,..,d+1~,
\end{align}
where the $z_i$ parametrize the sphere $S^{d-1}$, i.e.\ $\sum_{i}z_i^2=1$.
Euclidean AdS$_{d+1}$ as the interior of a unit ball $B^{d+1}$ is covered by these coordinates as follows. 
Sections of fixed time correspond to fixed latitude. 
The north/south poles correspond to $t\,{=}\,\pm\infty$.
The axis through them is $r=0$ and the surface of the ball with the two poles removed corresponds to $r=\infty$. 
Intermediate $r$ interpolate between these two extremes. 
The boundary at $r\,{=}\,\infty$ therefore is a cylinder $\mathbb{R}\,{\times}\, S^{d-1}$. 
Adding the two points corresponding to $t\,{=}\,\pm\infty$ completes the boundary to $S^{d-1}$. 
The resulting line element takes the form
\begin{equation}\label{eqn:metric-EAdS-global}
 ds^2=\big(L^2+r^2\big)d\tau^2+\frac{1}{1+r^2/L^2}dr^2+r^2 d\Omega_{d-1}^2~.
\end{equation}
The rescaling of the time coordinate as compared to the standard form of that metric is just for technical convenience.
The parametrizations (\ref{eqn:Poincare-coords}) and (\ref{eqn:EAdS-global}) can be combined to derive
the coordinate transformation connecting them.
We then straightforwardly find that the transformation corresponding to (\ref{eqn:WickRotationPoincare}) 
with $X_{0/1}\rightarrow \ii X_{0/1}=:\tilde X_{1/0}$ is in the coordinates (\ref{eqn:EAdS-global}) 
realized by simply setting
\begin{equation}\label{eqn:WickrotationsGlobal}
 H=\ii L~.
\end{equation}
With that analytic continuation the coordinatization of Euclidean AdS (\ref{eqn:EAdS-global}) 
becomes the dS parametrization
\begin{align}\label{eqn:dS-static-coords}
 \tilde X_0&=\sqrt{H^2-r^2}\sinh\tau~,&
 \tilde X_1&=\sqrt{H^2-r^2}\cosh\tau~,&
 X_i=r z_i\quad\forall i=2,..,d+1~.
\end{align}
The resulting line element is  -- up to a rescaling of the time coordinate -- that of the 
usual static-patch metric and reads
\begin{equation}\label{eqn:metric-ds-static-patch}
 ds^2=-\big(H^2-r^2\big)d\tau^2+\frac{1}{1-r^2/H^2}dr^2+r^2 d\Omega_{d-1}^2~.
\end{equation}
For  $r\,{\in}\,[0,H)$ (\ref{eqn:dS-static-coords}) parametrizes the static patch of dS which is thus related
to the inner shell $r\in[0,L)$ of Euclidean AdS, as illustrated in Fig.~\ref{fig:ds-ads1}. 
The transformation (\ref{eqn:WickrotationsGlobal}) then directly realizes the transformation discussed above 
in (\ref{eqn:WickRotationPoincare}) and used in \cite{Anninos:2011ui}.
It is obtained by simply transforming coordinates on both sides of that identification.
The relation of the dS static patch to only the inner shell of AdS reflects the fact that of the $\SO{1,d\,{+}\,1}$ isometries of 
dS$_{d+1}$ only an $\SO{d}\,{\times}\,\mathbb{R}$ subgroup, corresponding to the symmetries of $S^{d-1}$ and the timelike Killing 
field, preserves the horizon. Likewise, restricting to the inner shell of AdS also breaks the radial isometries.
The parametrization (\ref{eqn:dS-static-coords}) can also be continued to $r\,{>}\,H$, where $r$ becomes timelike and $t$ spacelike.
The roles of $X_0$ and $X_1$ are simply switched then, again by a double Wick rotation 
in the ambient space.
Only for $r\,{=}\,H$, corresponding to the horizon of the static patch, the 
transformation becomes singular.
The extension including $r\,{>}\,H$ relates almost all of the dS Poincar\'{e} patch
to global AdS. 
For definiteness we choose the expanding patch, such that the conformal boundary of 
AdS is mapped to the spacelike conformal boundary of dS at $\scri^+$.
The boundary arising if a finite cut-off is imposed on $r$ is timelike/spacelike so long as the cut-off is below/above the 
radius of curvature~$H$.

\section{Static patch holography from A\dS/CFT}\label{sec:staticpatch-holography}

Building on the higher-spin realization of dS/CFT via analytic continuation of AdS/CFT in \cite{Anninos:2011ui} and the 
identification of the static patch as part of AdS we will now attempt to realize static patch holography.
This will strongly build on the role of the AdS radial coordinate as an energy scale in the dual CFT.
Since many of the arguments do not depend on the spacetime dimension
we will mostly keep it general and only specialize to (A)dS$_4$ for certain points.

\subsection{From cut-off A\dS/CFT to cut-off \dS/CFT}\label{sec:cut-off-AdS-CFT-Wick}
In AdS/CFT cutting off the infrared part of the bulk geometry corresponds to a 
UV modification of the dual CFT and realizes a high-energy cut-off \cite{Susskind:1998dq,Peet:1998wn}.
In the semi-classical limit the values of on-shell bulk fields at fixed radial position 
can then be interpreted as running couplings in the dual CFT, and
the bulk field equations were related to RG equations of the boundary theory in \cite{deBoer:1999xf,*Balasubramanian:1999jd}.
Recently, more systematic approaches to a holographic realization of the Wilsonian renormalization group 
have been discussed in \cite{Faulkner:2010jy,Heemskerk:2010hk}.
These discussions employed AdS in Poincar\'{e} coordinates. 
To fix notation and set the stage, we now discuss the 
analog in the coordinates (\ref{eqn:EAdS-global}),~(\ref{eqn:metric-EAdS-global}).

We recall that AdS is conformally compact and choose a function $f=1/r$, 
defining the boundary via $\partial \mathcal M:=\lbrace p\,{\in}\,\mathcal M\,\vert\,f(p)\,{=}\,0\rbrace$.
The rescaled metric $\overline{g}:=f^2g$ then induces a representative of the boundary conformal structure 
on the conformal boundary. 
For explicitness we consider a bulk Klein-Gordon field $\phi$ of mass $m^2\,{=}\,-\Delta_+\Delta_-L^{-2}$ with $\Delta_\pm=d/2\pm\nu$ in the following,
but we expect similar results for fields of higher spin.
The asymptotic expansion of solutions is $\phi=f^{\Delta_-}\varphi^{}_-+f^{\Delta_+}\varphi^{}_+$ and we choose the 
standard quantization where the boundary-dominant part $\varphi_-$ is interpreted as source for the dual operator. 
The AdS/CFT prescription then reads
\begin{equation}\label{eqn:adscft-full}
 \mathcal Z[\varphi^{}_-]:=\int\mathcal D\phi\big\vert_{\phi\rightarrow f^{\Delta_-}\varphi^{}_-}\:e^{S}
 =\left\langle \exp \left\lbrace \int d^dx\sqrt{g^{}_\infty}\, \varphi^{}_-(x)\mathcal O(x)\right\rbrace\right\rangle_\text{CFT}~.
\end{equation}
On the right hand side $g^{}_\infty$ denotes the representative of the boundary conformal structure as explained above.
We then introduce a fixed value $r_\kappa:=\kappa$ for the radial coordinate and split the path integral into the parts 
corresponding to $r\leq r_\kappa$ and $r\geq r_\kappa$
\begin{align}\label{eqn:split-path-integral}
 \Psi^{}_\text{IR}[\phi^{}_\kappa]&:=\int^{\phi(r_\kappa)=f^{\Delta_-}\phi^{}_\kappa} \mathcal D\phi\big\vert^{}_{r\leq r_\kappa}\: e^{S+S_\kappa}~,&
 \Psi^{}_\text{UV}[\varphi^{}_-,\phi^{}_\kappa]&:=\int^{\phi\rightarrow f^{\Delta_-}\varphi^{}_-}_{\phi(r_\kappa)=f^{\Delta_-}\phi^{}_\kappa} \mathcal D\phi\big\vert^{}_{r\geq r_\kappa}\: e^{S-S_\kappa}~.
\end{align}
$S_\kappa\,{=}\,S_\kappa[\phi_\kappa]$ is an arbitrary boundary action at $r\,{=}\,r_\kappa$, which just introduces a multiplicative renormalization 
of both objects since it is fixed by the boundary condition at $r_\kappa$ and can be pulled out of the path integral.
We have normalized the boundary condition at $r_\kappa$ such that we can conveniently take the limit $\kappa\rightarrow \infty$.
The full path integral becomes
\begin{equation}\label{eqn:ZUVIR}
 \mathcal Z[\varphi^{}_-]= \int \mathcal D \phi^{}_\kappa\: \Psi^{}_\text{IR}[\phi^{}_\kappa]\, \Psi^{}_\text{UV}[\varphi^{}_-,\phi^{}_\kappa]~.
\end{equation}
Following \cite{Faulkner:2010jy,Heemskerk:2010hk} the generating function for correlators in the CFT with a cut-off at an energy scale $\Lambda_\kappa$ 
is then identified with $\Psi_\text{IR}$ by
\begin{equation}\label{eqn:AdS/cutoffCFT}
 \Psi^{}_\text{IR}[\phi^{}_\kappa]
 =\left\langle \exp\left\lbrace \int d^dx\sqrt{g_\kappa}\,\phi^{}_\kappa\mathcal O\right\rbrace\right\rangle_{\text{CFT},\Lambda_\kappa}~.
\end{equation}
Since (\ref{eqn:AdS/cutoffCFT}) restricts the bulk theory to a part of AdS, the \SO{1,d\,{+}\,1} bulk isometries 
are broken to those preserving the radial cut-off, $\mathbb{R}\,{\times}\,\SO{d}$.
In the dual theory that corresponds to the breaking of conformal invariance by the UV cut-off.
The analog in Poincar\'{e} coordinates preserves the boundary Euclidean symmetries $\SO{d}\ltimes\mathbb{R}^d$, 
and we recover the fact that while the conformal symmetries of the cylinder and the plane agree, their isometries do not.
The metric on the right hand side would naturally be that induced by  $\overline g$ at $r=r_\kappa$.
However, since the boost symmetries mixing the time and spatial directions are broken anyway, we can also use 
$g_{\kappa,\tau\tau}=(L^2+f^{-2})^{-1}g_{\tau\tau}$ and $g_{\kappa,ij}=f^2g_{ij}$ to extract the CFT metric.
Asymptotically that becomes equivalent to $\overline{g}$ and we recover the usual induced conformal structure.
For finite $\kappa$ this keeps the time component of the boundary metric normalized and ensures that, similarly 
to the prescription in Poincar\'{e} coordinates, changes in $\kappa$ have a purely field-theoretic interpretation\footnote{%
Alternatively, we can take as induced geometrical data on the boundary that appropriate for a non-relativistic theory,
i.e.\ a spatial metric along with an orthogonal timelike one-form, as discussed in  \cite{Ross:2011gu}.}.

\subsubsection*{Analytic continuation to dS}
Combining the above discussions with the arguments used in \cite{Anninos:2011ui} we can now derive
a holographic description for the static patch with radial cut-off as follows.
We start in the same way from  the GKPY duality relating the \O{N} CFT$_3$ to the minimal version 
of Vasiliev's higher-spin theory on AdS$_4$.
However, for the bulk AdS we employ the coordinates (\ref{eqn:EAdS-global}),(\ref{eqn:metric-EAdS-global}),
such that the dual CFT at $r\,{\rightarrow}\,\infty$ is defined on the cylinder.
We then use the same analytic continuation, which in our coordinates is realized by (\ref{eqn:WickrotationsGlobal}),
but apply it to the cut-off AdS/CFT prescription (\ref{eqn:AdS/cutoffCFT}) with $\kappa\,{<}\,L$.
That transforms the cut-off AdS bulk geometry to the corresponding part of the dS static patch with $r<\kappa$,
while the Euclidean CFT$_3$ on the cylinder is Wick-rotated to Lorentzian signature.
For the scalar discussed above it also switches the sign of the mass, in agreement with \cite{Iazeolla:2007wt}.
Following \cite{Anninos:2011ui} we note that, since $N\propto (\Lambda G_\text{N})^{-1}$ in the GKPY duality, this 
should on the CFT side be combined with $N\rightarrow -N$. A little more formally we obtain 
\begin{equation}\label{eqn:dS/cutoffCFT}
 \Psi_\text{IR}^\text{dS}[\phi^{}_\kappa]:=\Psi^{}_\text{IR}[\phi^{}_\kappa]\Big\vert_{L=\ii H}
 =\left\langle \exp\left\lbrace \int d^dx\sqrt{g_\kappa}\,\phi^{}_\kappa\mathcal O\right\rbrace\right\rangle_{\text{CFT},\Lambda_\kappa,\tau\rightarrow\ii\tau,N\rightarrow -N}~.
\end{equation}
This results in a duality of higher-spin gravity on a part of the dS static patch and a cut-off version of the 
\Sp{N} CFT$_3$ of \cite{Anninos:2011ui} on the Lorentzian-signature cylinder.
Since we have restricted to $\kappa$ smaller than the radius of curvature the bulk coordinates are regular on both, 
the dS and AdS sides of the analytic continuation. The cylinder as boundary geometry arises straightforwardly from the bulk
and we have a dual description for the static patch with a radial cut-off.
Note that this derivation is valid for the cut-off arbitrarily close to the horizon.
While the proposal is rather straightforward to implement from the bulk perspective, 
its interpretation on the CFT side poses some non-trivial questions.
In the AdS/CFT picture the radial direction is understood to be encoded in the CFT as RG flow
and thus has a clear interpretation. 
In the usual dS/CFT setting on the other hand, the time direction itself has to emerge 
in a non-trivial way from the CFT. 
It is not quite clear therefore, how cutting off e.g.\ the upper yellow-shaded triangle of the dS geometry 
in Fig.~\ref{fig:ds-ads1} to arrive at the static patch is reflected in the CFT at $\scri^+$.
A related issue is that the \Sp{N} CFT becomes non-unitary when Wick-rotated to Lorentzian signature. 
While such a continuation is not desired in \cite{Anninos:2011ui}, this is what happens with the 
cut-off version of the theory in (\ref{eqn:dS/cutoffCFT}).
How a UV cut-off can restore unitarity has been investigated in \cite{Andrade:2011aa}, 
and one could hope for a similar mechanism to be realized here.

\subsection{Holography for the static patch}\label{sec:holography-for-static-patch}
We now want to obtain a duality defined on the entire dS static patch.
To this end we have to consider the analytic continuation of the cut-off AdS/CFT duality (\ref{eqn:AdS/cutoffCFT})
to a cut-off static-patch holography (\ref{eqn:dS/cutoffCFT}) in the limit where $\kappa$ approaches the
radius of curvature.
There are no particular complications arising for the AdS bulk theory with cut-off at $r\,{=}\,L$ and it simply 
corresponds to the \O{N} CFT$_3$ on the cylinder with a particular value for the UV cut-off.
We could thus perform calculations on both sides of the AdS/CFT correspondence and then define the 
dS static patch/cut-off CFT picture by analytic continuation.
As discussed above, for $\kappa\,{<}\,L$ there is also no problem in the analytic continuation of the 
bulk$\leftrightarrow$boundary dictionary itself from (\ref{eqn:AdS/cutoffCFT}) to (\ref{eqn:dS/cutoffCFT}), 
to obtain a duality which is intrinsically defined on the dS static patch.

However, for the holographic dictionary intrinsically on the static patch the limit $\kappa\rightarrow H$ is 
non-trivial, due to the infinite red-shift factor in the bulk metric at the horizon.
As a result, only the $S^{d-1}$ part of the boundary cylinder $\mathbb{R}\times S^{d-1}$ arises naturally
from the bulk geometry in position space: sending $r$ to $H$ with the other coordinates fixed reduces the 
bulk geometry by two dimensions, as can be seen in Fig.~\ref{fig:ds-ads1a} from the fact that the 
constant-$\tau$ surfaces meet at a point on the horizon.
That challenges the interpretation of $\phi\vert_{r=H}$ as a source in the dual CFT and obscures the boundary
geometry.
It is therefore convenient for the formulation of a holographic dictionary for the entire static patch to
exploit the existence of a timelike Killing field and employ the Fourier transform, following \cite{Sachs:2001qb}.
For notational convenience we change the radial coordinate to $r\,{=}\,H\sech\frac{z}{H}$ such that the 
horizon $r\,{\rightarrow}\,H$ corresponds to $z\,{\rightarrow}\,0$.
As we shall verify in Sec.~\ref{sec:scalars-explicitly}, the asymptotic form of the bulk field as $z\rightarrow 0$ is given by
\begin{equation} \label{eqn:dS-horizon-expansion}
 \phi(\tau,z,x)=\int \frac{d\omega}{\sqrt{2\pi}}\: e^{-\ii \omega\tau}\left(\varphi^+_\omega(z,x) z^{\ii \omega}  +  \varphi^-_\omega(z,x) z^{-\ii \omega}\right)~,
\end{equation}
where $\varphi_\omega^\pm$ have regular power series expansions in $z$.
These correspond to the presence of left- and right-moving modes near the horizon\footnote{%
The Fourier modes in (\ref{eqn:dS-horizon-expansion}) become rapidly oscillating as the horizon is approached
and the Fourier transform becomes singular, since the timelike Killing field degenerates.},
and we have to adapt the Dirichlet boundary condition accordingly.
The situation is actually similar to a scalar field on AdS with mass below the Breitenlohner-Freedman stability 
bound, see App.~A of \cite{Andrade:2011nh}, and we derive admissible boundary conditions from the demand that we 
have to find a well-defined symplectic structure. 
We focus on the standard Klein-Gordon product defined from the
canonical symplectic current $j_\mu=\ii \phi_1\partialLR_\mu\phi_2$,
noting that other choices may be of interest as well \cite{Jafferis:2013qia}.
The flux through a surface of constant $r$ approaching the horizon is given by 
\begin{align}\label{eqn:horizon-flux}
 \mathcal F&=\int_{r\rightarrow H^-}\!d^dx\sqrt{-g_\text{ind}} n^\mu j_\mu
 =\int_{S^{d-1}}\int\! {d\omega}\:2\omega H^{d-1}\left(\varphi_{1,\omega}^+\varphi_{2,-\omega}^+-\varphi_{1,\omega}^-\varphi_{2,-\omega}^-\right)~,
\end{align}
where $n=\sqrt{g^{rr}}\partial_r$ is the unit normal vector field and the 
volume form in the integration over $S^{d-1}$ is implicit. 
Note that the vanishing volume form is compensated by the radial derivative combined with the oscillatory behavior of $\phi$, 
yielding a finite result.
Conservation then demands $\mathcal F=0$ and 
we thus find that the natural way to impose boundary conditions is on the oscillatory parts of the Fourier modes at the horizon.
Admissible boundary conditions for quantum fluctuations are for example given by demanding for all $\omega>0$
\begin{align}\label{eqn:static-patch-bc}
 (i)\ \delta\varphi_\omega^+&=\delta\varphi_{-\omega}^-=0  &  &\text{or}& (ii)\ \delta\varphi_\omega^-&=\delta\varphi_{-\omega}^+=0~.
\end{align}
How the entire bulk field can be reconstructed once the boundary values are fixed has been studied in \cite{Sachs:2001qb}.
We note that, with the horizon at $z=0$, $(i)$/$(ii)$ correspond to outgoing/ingoing boundary conditions, respectively.
The situation is similar to AdS fields close to the Breitenlohner-Freedman bound,
where two quantization prescriptions are available and either the boundary-dominant or 
sub-dominant component is identified as source for the dual operator \cite{Breitenlohner:1982bm, *Breitenlohner:1982jf,*Witten:2001ua}\footnote{%
In fact, with a radial cut-off on AdS Neumann and Dirichlet modes are normalizable independently of the mass.}.
More general mixed boundary conditions are possible in both cases.
As a book-keeping device we may shift $\omega\rightarrow\omega(1\pm\ii \epsilon)$ with $\epsilon\rightarrow 0^+$ understood.
The boundary condition (\ref{eqn:static-patch-bc}) then fixes the non-normalizable modes, and it is natural
to identify the corresponding boundary values as sources for gauge-invariant operators $\mathcal O_{\pm\omega}$ of the dual theory.
The dictionary (\ref{eqn:dS/cutoffCFT}) thus becomes
\begin{align}\label{eqn:staticdS-CFT}
  \Psi_\text{IR}^\text{dS}[\varphi_\omega^\pm,\varphi_{-\omega}^\mp]&=
  \left\langle \exp\left\lbrace 
  \ii \int_{S^{d-1}}\int_{\omega\geq 0} \!{d\omega}
 \left(\varphi_\omega^\pm\mathcal O_{-\omega}+\varphi_{-\omega}^\mp\mathcal O_\omega\right)
 \right\rbrace\right\rangle_{\text{CFT},\Lambda_\kappa}~,
\end{align}
where the upper choice of signs in $\varphi^\pm$/$\varphi^\mp$ corresponds to imposing the boundary 
condition $(i)$ and the lower choice to imposing $(ii)$.
We have just split $\lbrace\mathcal O_\omega, \omega\,{\in}\,\mathbb{R}\rbrace$ into 
$\lbrace (\mathcal O_{-\omega},\mathcal O_{\omega}$), $\omega\,{\geq}\,0\rbrace$ and re-assigned the sources.
Since we have identified the Fourier modes individually with dual operators building on the fact that the
Fourier transform becomes singular on the horizon, transforming back to position space in the dual theory
could be delicate.
The most conservative picture would be to understand the 
dual theory on $S^{d-1}$, with a family of operators labeled by $\omega$ that encodes the bulk time evolution.
However, the discussion of Sec.~\ref{sec:cut-off-AdS-CFT-Wick}, which was valid for the cut-off arbitrarily 
close to the horizon, strongly suggests that this data organizes into a cut-off CFT on the cylinder.
The boundary geometry just arises differently from the bulk:
the $S^{d-1}$ directly arises as holographic screen, while the $\mathbb{R}$ factor naturally 
arises in Fourier space.
The identification of the Dirichlet boundary condition to be imposed in (\ref{eqn:dS/cutoffCFT})
with those resulting from (\ref{eqn:horizon-flux}) seems non-trivial and deserves further investigation.
We will leave that for the future and for the time being note that (\ref{eqn:staticdS-CFT})
naturally realizes a concise holographic dictionary intrinsically on the dS static patch.

\subsection{The static patch entropy}\label{sec:entropy}
Associated to the cosmological horizon for the static-patch observer on dS
is a finite thermodynamic entropy \cite{Gibbons:1977mu}.
Much like for the general case of black holes
the microscopic origin of that entropy has remained elusive, in particular whether 
it is related to a counting of microstates in a quantum-gravitational description.
The dual description of higher-spin gravity on the static patch in terms of a cut-off CFT 
on the horizon provides a handle to gain some insight.
More concretely, we can derive the number of bulk degrees of freedom by counting those of 
the dual theory.

The identifications (\ref{eqn:dS/cutoffCFT}), (\ref{eqn:staticdS-CFT}) relate the 
static patch of dS with an optional radial cut-off to a dual QFT on the boundary.
Being defined on $S^2$ this boundary theory naturally has an IR cut-off. 
Moreover, it is the analytic continuation of a boundary theory on AdS where the bulk has a radial cut-off.
The boundary theory therefore also has a cut-off in the UV,
and we expect a finite number of degrees of freedom. 
To make this more precise
we start with the cut-off AdS/CFT picture (\ref{eqn:AdS/cutoffCFT}) and repeat 
the analysis of \cite{Susskind:1998dq} for the higher-spin theory in the bulk AdS$_4$ and 
the \O{N} CFT$_3$ on the $\mathbb{R}\,{\times}\,S^2$ boundary.
A convenient way to implement the UV cut-off corresponding to the bulk IR cut-off 
$r\,{\leq}\,r_\kappa$ 
in the boundary theory is by introducing a minimal length and discretizing the 
boundary geometry, to obtain a lattice. 
Defining a dimensionless parameter by $r_\kappa=:L\delta^{-1}$, such that
$\delta\,{\rightarrow}\,0$ corresponds to full AdS, 
the $S^2$ is then naturally composed of $\mathcal O(\delta^{-2})$ 
cells, with each boundary field having one degree of freedom per cell. 
The overall coefficient depends on the specific realization of the UV cut-off and shall
not bother us.
The vector-like boundary theories we are dealing with only have $\mathcal O(N)$ degrees of freedom, 
as compared to $\mathcal O(N^2)$ in the usual Yang-Mills theories. 
We thus find the total number $n$ of degrees of freedom in the boundary theory
$n\,{\propto}\,N \delta^{-2}$.
A more geometric meaning can be given to the bulk radial cut-off
by noting that the surface area  of the cut-off AdS is $A_\kappa\,{\propto}\,r_\kappa^2= L^2\delta^{-2}$.
We thus find  $n\,{\propto}\, N A_\kappa L^{-2}$.
As noted in Sec.~\ref{sec:cut-off-AdS-CFT-Wick} we have $N\,{\propto}\,(\Lambda G_\text{N})^{-1}$ in the GKPY duality, 
where $G_\text{N}$ and $\Lambda$ are Newton's and the cosmological constant in the bulk, respectively.
Combining that with $L\,{\propto}\, \Lambda^{-1/2}$ in the four-dimensional bulk theory, we arrive at 
\begin{equation}
 n\propto \frac{A_\kappa}{G_\text{N}}~.
\end{equation}
This is the number of degrees of freedom for the cut-off CFT in the AdS/CFT picture (\ref{eqn:AdS/cutoffCFT}). 
With the GKPY duality we have thus obtained that the higher-spin bulk theory on AdS respects a holographic bound 
of the form discussed in \cite{Susskind:1994vu}.

The analytic continuation in (\ref{eqn:dS/cutoffCFT}) does not change the number of degrees of freedom of the boundary
theories, and we can thus transfer this result to our static-patch/cut-off-CFT duality: the dual description
of the static patch also has $n\,{\propto}\, A_\kappa/G_\text{N}$ degrees of freedom.
Deriving the corresponding entropy is particularly simple for the boundary theory with anticommuting scalars.
In a Fock-space representation the occupancy of each degree of freedom is at most one, such 
that the dimension of the Hilbert space $\mathcal H$ is $2^n$. For the entropy we thus find
\begin{equation}\label{eqn:entropy}
 \mathcal S=\log\operatorname{dim}\mathcal H\propto\frac{A_\kappa}{G_\text{N}}~.
\end{equation}
For the specific case that we holographically describe the entire static patch 
the cut-off is at $r_\kappa\,{=}\,H$ and $A_\kappa\,{=}\,4\pi H^2$.
Up to the undetermined overall coefficient (\ref{eqn:entropy}) then reproduces the horizon entropy.
Note that as $H\,{\rightarrow}\,\infty$, where flat space is recovered, also 
$n\rightarrow\infty$ and we correctly find an infinite entropy.

\section{Scalar fields explicitly}\label{sec:scalars-explicitly}
In this part we specialize to a free bulk scalar field and explicitly verify the transformation
from cut-off AdS/CFT to (cut-off) static-patch holography for the two-point functions of the dual operators.
We will be particularly interested in the entire static patch as bulk geometry, 
for which we recover the quasinormal frequencies from the dual theory on the horizon.
Although the scalar of Vasiliev's minimal higher-spin theory has $m^2L^2=-2$ and $d=3$
we will keep the mass and spacetime dimension general.
We also find it convenient to work with an action reproducing the scalar field equations,
although that may not be available for the full higher-spin theory.
For Euclidean AdS with the metric (\ref{eqn:metric-EAdS-global}) we start from
\begin{equation}\label{eqn:KGaction-AdS}
 S_\text{AdS}=-\frac{1}{2}\int d^{d+1}x\sqrt{g}\Big(g^{\mu\nu}\partial_\mu\phi \partial_\nu\phi-\mu^2L^{-2}\phi^2+V[\phi]\Big)~,
\end{equation}
such that the usual parametrization of the mass translates to $\mu^2=\Delta_+\Delta_-$.
Performing the analytic continuation to dS via (\ref{eqn:WickrotationsGlobal}) we note that the mass term switches
the sign.
That realizes the analytic continuation in the higher-spin theory \cite{Iazeolla:2007wt} and leaves $\Delta_\pm$ unchanged.
This is in fact also necessary to make sense of the boundary conditions in (\ref{eqn:dS/cutoffCFT}) for all values of the radial cut-off.
The resulting dS action reads
\begin{equation}\label{eqn:KGaction-dS}
 S_\text{dS}=-\frac{1}{2}\ii\int d^{d+1}x\sqrt{-g}\Big(g^{\mu\nu}\partial_\mu\phi \partial_\nu\phi+\mu^2H^{-2}\phi^2+V[\phi]\Big)~,
\end{equation}
where the metric is that of (\ref{eqn:metric-ds-static-patch}).
To have a positive mass on dS admissible $\nu$ are thus restricted to 
$0\,{\leq}\,\nu\,{\leq}\,d/2$ and we end up in the complementary series on dS\footnote{%
A snapshot review and references can be found in \cite{Andrade:2011nh}.
The fundamental series corresponds to imaginary $\nu$, and such fields on
AdS were also discussed briefly in \cite{Andrade:2011nh}.}.
For holographic applications the bulk actions (\ref{eqn:KGaction-AdS}), (\ref{eqn:KGaction-dS}) have to be renormalized 
in the usual way by adding counterterms at the conformal boundaries to render the combination finite on shell.
In the following we will simply drop contact terms in the correlators without explicitly constructing the counterterms.
For notational convenience we introduce $\lambda$ which is defined by $\lambda=L$ 
on AdS and $\lambda=\ii H$ on dS.
The Klein-Gordon equation on (A)dS resulting from (\ref{eqn:KGaction-AdS}), (\ref{eqn:KGaction-dS}) then reads 
\begin{align}\label{eqn:AdS-KG-Laplacian}
(\square+\mu^2\lambda^{-2})\phi&=0~,& \square&=r^{1-d}\partial_r(1+{r^2}/{\lambda^2})r^{d-1}\partial_r+\frac{\partial_\tau^2}{\lambda^2+{r^2}}+r^{-2}\bigtriangleup_{\text{S}^{d-1}}~.
\end{align}
To exploit the symmetries for its solution we employ in both cases the Fourier ansatz 
\begin{equation}
\phi\,{=}\,e^{-\ii\omega \tau}\, Y_{\vec{\ell}\,}(\Omega_{d-1})\,\chi^{}_\ell(r)~,
\end{equation}
with the spherical harmonics 
satisfying $\bigtriangleup_{\text{S}^{d-1}}Y_{\vec{\ell}}=-\ell(\ell+d-2)Y_{\vec{\ell}}$.
With $\chi^{}_\ell\,{=}\,(1-u)^{\Delta_+/2}u^{\ell/2}h(u)$, where 
$u\,{=}\,1/(1+\lambda^2/r^2)$,
the Klein-Gordon equation then 
translates to a hypergeometric differential equation for $h$.
The origin of AdS corresponds to $u\,{=}\,0$ approached from above and the position of 
the static-patch observer on dS to $u\,{=}\,0$ approached from below.
The solution which is normalizable at $r\,{=}\,0$ is thus in both cases given by
\begin{align}\label{eqn:ads-IR-radial-mode}
 \chi^{}_\ell(r,\omega)&=C_\kappa u^{\ell/2} (1-u)^{\Delta_+ /2} \, _2F_1\Big(\delta_+,\delta_-;\frac{d}{2}+\ell;u\Big)~,
 & \delta_\pm&=\frac{1}{2}(\Delta_++\ell\pm \ii \omega)~.
\end{align}
Note that $\chi^{}_\ell$ is real for appropriately chosen $C_\kappa$ and $\chi^{}_\ell(r,\omega)=\chi^{}_\ell(r,-\omega)$.
Furthermore, since the definitions of $u$ on dS and AdS are related by (\ref{eqn:WickrotationsGlobal}) so are the solutions.

\subsection{Two-point functions in cut-off (A)\dS/CFT}\label{sec:scalars-explicitly-cutoff-AdS}
As discussed in \cite{Balasubramanian:2012hb} it is convenient to employ for the split path integral (\ref{eqn:ZUVIR})
a subtraction scheme where the boundary action $S_\kappa$ in (\ref{eqn:split-path-integral}) coincides with the 
UV counterterms at the conformal boundary in the limit $\kappa\rightarrow\infty$, 
i.e.\ $S_\kappa\,{=}\,-\frac{1}{2}\int_{r=\kappa}\lambda^{-1}\Delta_-\phi^2+\dots$.
This ensures that $\Psi_\text{IR}$ becomes the partition function of the full theory as $\kappa\rightarrow \infty$.
In the semi-classical limit the inner part of the bulk path integral reduces to $\Psi_\text{IR}=e^{S_\text{IR}}$, where
\begin{equation}\label{eqn:Sir}
 S_\text{IR}=-\frac{1}{2}\int_{r=\kappa}d^dx\sqrt{g_\text{ind}}\:\phi\sqrt{g^{rr}}\partial_r\phi\:+\:S_\kappa~.
\end{equation}
The volume form becomes imaginary for dS and reproduces the overall factor $\ii$ in (\ref{eqn:KGaction-dS}).
The constant $C_\kappa$ in (\ref{eqn:ads-IR-radial-mode}) has to be chosen appropriately to satisfy the boundary condition at $r_\kappa$, 
e.g.\ for the Dirichlet boundary condition normalized as in (\ref{eqn:split-path-integral}) it
follows from $\chi^{}_\ell\vert^{}_{r=\kappa}\,{=}\,f^{\Delta_-}$.
The bulk solution with the boundary condition $\phi\vert^{}_{r\,{=}\,\kappa}=f^{\Delta_-}\phi_\kappa$ then reads
\begin{equation}\label{eqn:solphiads}
 \phi=\sum_{\vec{\ell}}\int \frac{d\omega}{\sqrt{2\pi}}\,e^{-\ii\omega \tau}\,\tilde\phi^{}_{\kappa,\vec{\ell}\,}(\omega)\,Y_{\vec{\ell}\,}(\Omega)\,\chi^{}_\ell(r,\omega)~,
\end{equation}
where $\tilde\phi_{\kappa,\vec{\ell}\,}(\omega)$ are the Fourier modes of $\phi_\kappa$ on $\mathbb{R}\,{\times}\,$S$^{d-1}$.
Inserting  (\ref{eqn:solphiads}) into (\ref{eqn:Sir}) 
we arrive at
\begin{align}\label{eqn:Sir2}
 S_\text{IR}&=\frac{1}{2}\:\sum_{\vec{\ell}}\int_{r=\kappa} \!{d\omega}\: \tilde\phi^{}_{\kappa,\vec{\ell}\,}(\omega)\:
             G_{\kappa,\ell}(\omega) \tilde\phi^{}_{\kappa,\vec{\ell}\,}(-\omega)~,
\end{align}
where
$G_{\kappa,\ell}(\omega)=r^{d-1}\sqrt{1+r^{2}/\lambda^{2}}\:\chi^{}_\ell(r,\omega)\big[\sqrt{\lambda^2+r^2}\partial_r \chi^{}_\ell(r,\omega)+\Delta_-\chi^{}_{\ell}(r,\omega)\big]$.
The cut-off CFT two-point functions are then obtained from the Fourier transforms of (\ref{eqn:AdS/cutoffCFT}) and (\ref{eqn:dS/cutoffCFT}).
Similarly to the transformation from (\ref{eqn:KGaction-AdS}) to (\ref{eqn:KGaction-dS}) 
via (\ref{eqn:WickrotationsGlobal}), the dS version picks up a factor $\ii$, which we absorb in $c_\lambda=e^{ \ii \arg\lambda}$.
We then arrive at 
\begin{equation}\label{eqn:AdS/cutoffCFT-FT}
 \Psi_\text{IR}^\mathrm{(A)dS}[\tilde\phi^{}_{\kappa,\vec{\ell}\,}]
 =\Big\langle \exp\Big\lbrace  -c_\lambda\:\sum_{\vec{\ell}}\int  \!{d\omega}\: \tilde\phi^{}_{\kappa,\vec{\ell}\,}(\omega)\tilde{\mathcal O}_{\vec{\ell}\,}(-\omega)\Big\rbrace\Big\rangle_{\text{CFT},\Lambda_\kappa}~,
\end{equation}
which yields 
$\big\langle\tilde{\mathcal O}_{\vec{\ell}_1\,}(\omega_1) \tilde{\mathcal O}_{\vec{\ell}_2\,}(\omega_2)\big\rangle_{\text{CFT},\Lambda_\kappa}{=}\,
 \delta_{\vec{\ell}_1\vec{\ell}_2}\delta(\omega_1+\omega_2)\,c_\lambda^{-2}\,G_{\kappa,\ell_1}(\omega_1)$.
Evaluating $G_{\kappa,\ell}(\omega)$ at $r\,{=}\,\kappa$ using standard identities for hypergeometric functions as found e.g.\ in \cite{Olver:2010:NHMF} 
and dropping contact terms results in
\begin{align}\label{eqn:cylinder-greens-function-G-cutoff}
 G_{\kappa,\ell}(\omega)&=2\,C_\kappa^2 \,\lambda^{d-1} \: u_\kappa^{\ell+d/2}(1-u_\kappa)^{\nu} X(\delta_+,\delta_-,\frac{d}{2}+\ell,u_\kappa)~,
\end{align}
where $u_\kappa\,{=}\,1/(1\,{+}\,\lambda^2/\kappa^2)$ and
$X(a,b,c,x)\,{=}\,c^{-1}(c-a)(c-b){}_2F_1(a,b,c+1,x){}_2F_1(a,b,c,x)$.
We have thus obtained the two-point functions of the dual cut-off CFTs on dS and AdS, 
with the choice of the constant $C_\kappa$ encoding the choice of boundary condition on the cut-off surface.
The parallel calculations leading to (\ref{eqn:cylinder-greens-function-G-cutoff}) establish their 
relation by the analytic continuation (\ref{eqn:WickrotationsGlobal}).
As compared to \cite{Anninos:2011ui} the continuation does not just affect the overall normalization,
reflecting the fact that the bulk radial cut-offs have different but apparently still related interpretations in the 
boundary theories on AdS and dS.
The two-point function (\ref{eqn:cylinder-greens-function-G-cutoff}) obtained from AdS with $\lambda\,{=}\,L$
has no poles for real $\omega$, as expected for the Euclidean boundary theory. 
The same applies for dS with $\kappa\,{>}\,H$, where the cut-off surface is spacelike and the boundary theory Euclidean.
For smaller $\kappa$ and $\lambda=\ii H$ the poles appear for real $\omega$, as expected for the Lorentzian boundary theory
on the dS static patch. 
The analytic continuation $L\rightarrow\ii H$ then yields the $\ii\epsilon$-prescription corresponding to the Wick rotation 
on the right hand side of (\ref{eqn:dS/cutoffCFT}).

\begin{figure}[ht]
\center
  \includegraphics[width=0.5\linewidth]{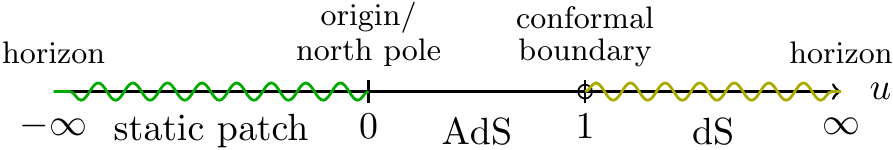}
\caption{The ranges of $u$ corresponding to the different parts of (A)dS.
         The colors for the dS parts correspond to those in Fig.~\ref{fig:ds-ads1a}.
         To avoid the branch points of the hypergeometric function the analytic continuation
         from $\lambda=L$ to $\lambda=\ii H$ should proceed through finite non-zero values, 
         e.g.\ via $\lambda=(1-t)L+\ii t H$.
         \label{fig:u-coord}}
\end{figure}
To complete the discussion we now consider the limit $\kappa\,{\rightarrow}\,\infty$.
The cut-off surface then approaches $\scri^+$ on dS and correspondingly the conformal
boundary of AdS, such that $\Psi^\mathrm{(A)dS}_\text{IR}$ becomes the full path integral.
On AdS as well as on dS $\kappa\rightarrow\infty$ corresponds to $u_\kappa\rightarrow 1$, 
see Fig.~\ref{fig:u-coord}, and the asymptotic expansions coincide.
With $\lim_{\kappa\rightarrow \infty} C_\kappa ={\Gamma(\delta_+)\Gamma(\delta_-)}/(\lambda^{\Delta_-}c_\lambda^{2\nu}{\Gamma(\nu)\Gamma(d/2+\ell)})$ we find
\begin{equation}\label{eqn:cylinder-greens-function-G}
 \lim_{\kappa\rightarrow \infty}G_{\kappa,\ell}(\omega)=2\lambda^{2\nu-1}c_\lambda^{-2\nu} \frac{\Gamma(1-\nu)\Gamma(\delta_+)\Gamma(\delta_-)}{\Gamma(\nu)\Gamma(\delta_+-\nu)\Gamma(\delta_--\nu)}~.
\end{equation}
The dependence on $\kappa$ has dropped out and the effect of the analytic continuation (\ref{eqn:WickrotationsGlobal}) 
is now restricted to the overall normalization, in accordance with the discussion of \cite{Anninos:2011ui}.
As a CFT two-point function (\ref{eqn:cylinder-greens-function-G}) should be conformally covariant, restricting its 
form to a power of the invariant distance of the two points on the cylinder.
Transforming back to position space for (A)dS$_2$ we indeed find
$\left\langle \mathcal O(0)\mathcal O(t,x)\right\rangle\,{\propto}\,\left(\sinh\frac{\tau}{2}\right)^{-2\Delta_+}$.
The poles and zeros in (\ref{eqn:cylinder-greens-function-G}) also encode the \mbox{(anti-)}quasinormal frequencies 
of dS$_{d+1}$, $\pm\ii\omega\,{=}\, 2n+\ell+\Delta_\pm$,
which have been calculated in~\cite{LopezOrtega:2006my}.
That they can be recovered 
from the dual CFT at $\scri^+$ in the dS/CFT proposal of \cite{Strominger:2001pn} has been emphasized already in~\cite{Abdalla:2002hg}.
A quantization prescription based on the quasinormal modes can be found in \cite{Jafferis:2013qia}.

\subsection{Static patch holography and quasinormal frequencies}\label{sec:scalar-explicitly-staticpatch}
We now specialize to the entire static patch which is recovered for $\kappa\,{\rightarrow}\, H^-$. The boundary data corresponding to sources for dual operators 
was naturally identified in Fourier space and we calculate the two-point functions using the dictionary (\ref{eqn:staticdS-CFT}).
Note first of all that the asymptotic expansion of the radial mode (\ref{eqn:ads-IR-radial-mode}) around the 
horizon at $u\,{\rightarrow}\,{-}\,\infty$ confirms (\ref{eqn:dS-horizon-expansion}), which we used to
derive the dictionary.
To actually calculate the left hand side of (\ref{eqn:staticdS-CFT}) in the saddle point approximation 
we have to evaluate 
$\Psi_\text{IR}^\text{dS}[\varphi_\omega^\pm,\varphi_{-\omega}^\mp]=e^{S^{\mathrm{dS}}_\mathrm{IR}}$. 
Evaluating the action (\ref{eqn:KGaction-dS}) on shell with the 
expansion (\ref{eqn:dS-horizon-expansion}) yields
\begin{align}\label{eqn:KGaction2onshell}
  S^\mathrm{dS}_{\mathrm{IR},\kappa\rightarrow H^-}&= -H^{d-1}\int_{S^{d-1}}\int_{\omega\geq 0}\!{d\omega}\:
  \underbrace{\frac{1}{2}\ii\left[ \tilde\phi(\omega)z\partial_z\tilde\phi(-\omega)-\tilde\phi(-\omega)z\partial_z\tilde\phi(\omega)\right]_{z=0}}_{=\omega\left(\varphi_{\omega}^+\varphi_{-\omega}^+-\varphi_{\omega}^-\varphi_{-\omega}^-\right)}~.
\end{align}
For fixed $\omega>0$ and the boundary condition $(i)$ in (\ref{eqn:static-patch-bc}) the functions 
$\varphi_\omega^{+}$ and $\varphi_{-\omega}^{+}$ evaluated at $z\,{=}\,0$ thus constitute
a pair of source and expectation value for the dual operator $\mathcal O_{-\omega}$, 
while $\varphi_{-\omega}^{-}$ and $\varphi_{\omega}^{-}$ are source and expectation value for $\mathcal O_{\omega}$, respectively.
For the alternative quantization with the boundary condition $(ii)$ the analogous statement applies
with $\varphi^+$ and $\varphi^-$ exchanged.
For the two-point functions  we then find
$\big\langle \mathcal O_{\omega_1,\vec{\ell}_1}\mathcal O_{\omega_2,\vec{\ell}_2}\big\rangle
=-\delta_{\vec{\ell}_1\vec{\ell}_2}\delta(\omega_1+\omega_2)\,G_{\omega_1,\ell_1}$, where
\begin{align}
 G^{(i)}_{\omega,\ell}&=
 \frac{\delta}{\delta\varphi^-_{-\omega}}\frac{\delta}{\delta\varphi^+_{\omega}}\:e_{}^{S^\mathrm{dS}_{\mathrm{IR},\kappa\rightarrow H^-}}
 =
 2\omega H^{d-1}\frac{\delta\varphi_{\omega}^-}{\delta\varphi_{\omega}^+}~,
 &G^{(ii)}_{\omega,\ell}&=
 -2\omega H^{d-1}\frac{\delta\varphi_{\omega}^+}{\delta\varphi_{\omega}^-}~.
\end{align}
They are therefore almost reciprocal to each other.
The precise forms are most conveniently derived from the asymptotic expansion of 
the full bulk solution around $z=0$.
To implement the boundary condition on the horizon we have to fix
$C_\kappa^{(i)}=C_\kappa^{(ii)}\vert_{\omega\rightarrow-\omega}=\frac{H^{\ii\omega} \Gamma \left(\delta _-\right) \Gamma \left(\delta _--\nu \right)}{\ii^{\ell}\Gamma (-\ii \omega ) \Gamma \left(d/2+\ell\right)}$ 
in (\ref{eqn:ads-IR-radial-mode}), and the bulk field is then again given by (\ref{eqn:solphiads}).
From the expansion of (\ref{eqn:ads-IR-radial-mode}) with $u=-\operatorname{csch}^2(z/H)$ we then find
\begin{align}\label{eqn:horizon-two-point}
 G^{(i)}_{\omega,\ell}&=G^{(ii)}_{-\omega,\ell}=-
 2\ii H^{d+2\ii\omega-1}
  \frac{\Gamma(1+\ii \omega)}{\Gamma(-\ii \omega)}
  \frac{\Gamma(\delta^{}_-)\Gamma(\delta^{}_--\nu)}{\Gamma(\delta^{}_+)\Gamma(\delta^{}_+-\nu)}~.
\end{align}
The boundary condition on AdS corresponding to the choice of $C_\kappa$ is not directly Dirichlet,
but since the bulk solution is fixed uniquely the two can be related.
With the choices naturally appearing at the horizon, $C_\kappa=C_\kappa^{(i)/(ii)}$, we could have obtained (\ref{eqn:horizon-two-point}) 
up to contact terms also from the near-horizon limit of (\ref{eqn:cylinder-greens-function-G-cutoff}).
The frequency dependence of the right hand side of (\ref{eqn:horizon-two-point}) in particular encodes the dS$_{d+1}$ quasinormal frequencies: we find
poles and zeros for
\begin{align}
 \ii\omega&=\Delta_\pm+\ell+2n & \text{and}&& -\ii\omega&=\Delta_\pm+\ell+2n~,
\end{align}
respectively, where $n\,{\in}\,\mathbb{N}$.
These are precisely the (anti-)quasinormal frequencies arising for scalar perturbations as found in \cite{LopezOrtega:2006my}.
Since they are naturally associated to the static patch of dS$_{d+1}$ it is desirable to recover them 
from a dual description which is intrinsically defined on the static patch.
Previously they have been recovered from a putative quantum mechanics on the observer worldline in \cite{Anninos:2011ui},
and here we find them encoded directly on the horizon of the static patch as the natural place to define the dual theory.

\section{Incorporating the {\dS/\dS} correspondence}\label{sec:dsds}

In this section we come back to a seemingly disconnected proposal for a holographic description of dS:
building on the complementary holographic interpretation of Randall-Sundrum 
setups \cite{PhysRevLett.85.2052,*Giddings:2000mu} and warped geometries where dS$_{d+1}$ 
is sliced by dS$_d$, a dual description for dS$_{d+1}$ gravity in terms of 
cut-off CFTs on dS$_d$ was proposed in \cite{Alishahiha:2004md,Alishahiha:2005dj}.
The discussion of static-patch holography above can nicely be adapted to this dS/dS correspondence, 
and then similarly allows to lift it to a concrete level. 
We can thus incorporate the dS/dS correspondence into a coherent picture of dS holography via cut-off AdS/CFT.
In the first part of the section we derive a relation of the dS/dS geometry to a part of AdS,
similarly to the discussion of Sec.~\ref{sec:dS-AdS-relation}.
Building on that identification we can then realize analytic continuations from AdS/CFT to 
dS/dS correspondences.

\begin{figure}[ht]
\center
\subfigure[][]{ \label{fig:dsds-global}
  \includegraphics[width=0.34\linewidth]{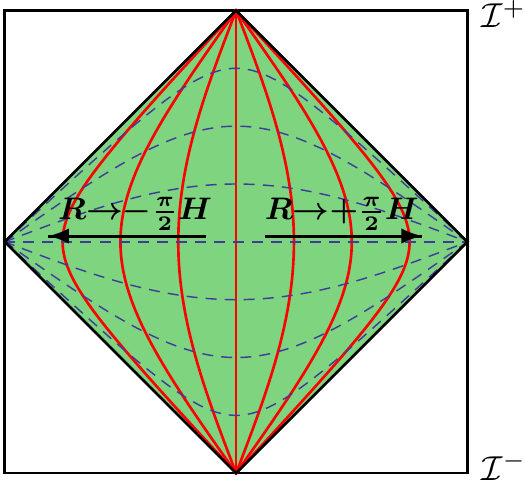}
}\qquad\qquad\quad
\subfigure[][]{ \label{fig:adsads}
    \includegraphics[width=0.31\linewidth]{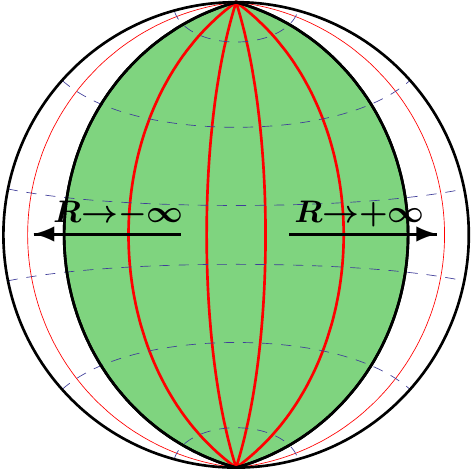}
}
\caption{The left hand side shows the slicing of dS$_{d+1}$ by dS$_d$ hypersurfaces.
         On the horizontal axis is the polar angle on S$^d$ and the boundaries correspond to the north and south pole.
         The red solid and blue dashed curves correspond to constant $R$ and global dS$_d$ time, respectively.
         The right hand side similarly shows the slicing of AdS$_{d+1}$ by AdS$_d$ hypersurfaces.
         The green shaded regions are identified by the analytic continuation (\ref{eqn:wick-dsds-adsads}).
         Note that, with a cut-off on $R$, also on the AdS$_{d+1}$ side only an S$^{d-1}$ of 
         the S$^d$ conformal boundary is covered.}         
\end{figure}

The geometries for the dS/dS correspondence are obtained from the fact that,
with a parametrization $x_i$ of dS$_d$ as hyperboloid with radius of 
curvature $h$ in $\mathbb{R}^{1,d}$,
one obtains a parametrization of dS$_{d+1}$ via
\begin{align}\label{eqn:dsds-embedding}
 X_k&=\frac{H}{h}\cos\frac{R}{H}\:x_k~,\quad\forall k=0,..,d~,& X_{d+1}&=H\sin\frac{R}{H}~.
\end{align}
From $-x_0^2+\sum_{i=1}^{d}x_i^2=h^2$ we find that these coordinates
cover some part of dS$_{d+1}$ with radius of curvature $H$ as in (\ref{eqn:hyperboloids}).
The part which is covered comprises all of the spatial S$^d$ factor for $X_0\,{=}\,0$ 
and shrinks to an S$^{d-1}$ subspace at $\scri^\pm$, see Fig.~\ref{fig:dsds-global}.
To actually fix the geometry we have yet to specify which part of dS$_d$ the slices cover.
We may choose global dS$_d$ slices, for which the above discussion applies, or restrict e.g.\ to the 
expanding Poincar\'{e} patch as illustrated for dS$_{d+1}$ in Fig.~\ref{fig:ds-ads1a}.
That choice covers the $\mathbb{Z}_2$ quotient of dS$_d$ relevant for the `elliptic interpretation' 
going back to \cite{schroedinger1956expanding}.
A third option is to choose only the static patches of the dS$_d$ slices, and we will 
refer to the corresponding dS/dS geometries as global, elliptic and static dS/dS.
Our focus for deriving dS/dS correspondences will be on the elliptic and static dS/dS geometries, for which a correspondingly
smaller part of dS$_{d+1}$ is covered.
More precisely, the Poincar\'{e} coordinates cover only the part $x_0>x_1$ of the slices.
Via (\ref{eqn:dsds-embedding}) this implies also $X_0>X_1$,
and the elliptic dS/dS geometry thus is a part of the Poincar\'{e} patch of dS$_{d+1}$ illustrated in Fig.~\ref{fig:ds-ads1a}.
We can therefore employ the analytic continuations used in \cite{Anninos:2011ui} and similarly to the construction in Sec.~\ref{sec:staticpatch-holography}
restrict to the region appropriate for the dS/dS geometry.
The same applies for static dS/dS. 
In that case the bulk geometry is manifestly static and the coordinates indeed cover the dS$_{d+1}$ static patch.
The resulting line element in any case reads
\begin{align}
 ds^2_{d+1}&=dR^2+\frac{H^2}{h^2}\cos^2\frac{R}{H}\:ds^2_{\mathrm{dS}_d}~.
\end{align}
To obtain an analytic continuation to an AdS geometry we employ the Wick rotations discussed
in Sec.~\ref{sec:dS-AdS-relation} and apply it to the slices, e.g.\ the analog of (\ref{eqn:WickRotationPoincare}) 
for the slices of elliptic dS/dS, or similarly (\ref{eqn:WickrotationsGlobal}) for static-patch slices.
We then perform in (\ref{eqn:dsds-embedding}) the analytic continuation
\begin{align}\label{eqn:wick-dsds-adsads}
 H&\rightarrow \ii L~,&  h&\rightarrow\ii l~, & &\dots
\end{align}
where the dots denote the possible further transformations to complete $h\rightarrow \ii l$ to an analytic continuation
of the dS$_{d}$ slices to AdS$_{d}$.
The parametrization (\ref{eqn:dsds-embedding}) becomes
\begin{align}\label{eqn:adsads-embedding}
 X_k&=\frac{L}{l}\cosh\frac{R}{L}\:x_k~,\quad\forall k=0,..,d~,& X_{d+1}&=L\sinh\frac{R}{L}~,
\end{align}
and since $-X_0^2+\sum_i X_i^2=-L^2$ we find a parametrization of a part of AdS$_{d+1}$.
The line element becomes
\begin{align}
 ds^2_{d+1}&=dR^2+\frac{L^2}{l^2}\cosh^2\frac{R}{L}\:ds^2_{\mathrm{AdS}_d}~,
\end{align}
and that geometry is illustrated in Fig.~\ref{fig:adsads}.
The complete AdS$_{d+1}$ corresponds to $R\,{\in}\,\mathbb{R}$. 
The analytic continuation
(\ref{eqn:wick-dsds-adsads}) relates the dS$_d$ slicing  of dS$_{d+1}$ to an inner part of the AdS$_d$ slicing of AdS$_{d+1}$, 
where the radial coordinate is restricted to $|R|\leq {\pi L}/{2}$.
For elliptic dS/dS the analytic continuation yields complete AdS$_d$ slices,
while for static dS/dS it introduces a cut-off not only on the AdS$_{d+1}$ radial
coordinate $R$ but also on the radial coordinate of the AdS$_d$ slices.

We have thus obtained for the elliptic and static dS/dS geometries a geometric identification with 
a part of AdS, and can proceed to implement a similar analytic continuation of the AdS/CFT dictionary as 
we have done for the static patch in (\ref{eqn:dS/cutoffCFT}).
For static dS/dS the spatial cut-off on the AdS$_d$ slices of the corresponding AdS$_{d+1}$ geometry 
calls for a separate treatment, and we therefore start with a discussion of elliptic dS/dS.
In the AdS/CFT picture the conformal boundary of AdS$_{d+1}$
comprises two copies of AdS$_{d}$ joined at their conformal boundaries, as illustrated in Fig.~\ref{fig:adsads}.
The bulk theory is thus dual to a pair of CFTs on AdS.
One may study them with transparent boundary conditions or identify them by considering AdS$_{d+1}$/$\mathbb{Z}_2$,
with the slices at $\pm R$ identified, as in \cite{Aharony:2010ay,Andrade:2011nh}.
Each choice leads to a distinct AdS/CFT duality, and we keep that general in the following.
Introducing a radial cut-off $|R|\,{\leq}\,\pi\kappa/2\,{=:}\,R_\kappa$ on AdS would again be interpreted as a UV cut-off in the dual CFTs,
and the analog of the cut-off AdS/CFT duality (\ref{eqn:AdS/cutoffCFT}) then reads
\begin{equation}\label{eqn:AdSAdS/cutoffCFT}
 \Psi_\text{IR}[\phi^{+}_{\kappa},\phi^{-}_{\kappa}]
 =\left\langle \exp\left\lbrace \int_{\mathrm{AdS}_d}\! \big(\phi^{+}_{\kappa}\,\mathcal O^{}_+ +\phi^{-}_{\kappa}\,\mathcal O^{}_-\big)\right\rbrace\right\rangle_{\text{CFT},\Lambda_\kappa}~.
\end{equation}
The inner part of the bulk path integral $\Psi_\text{IR}$ is defined analogously to (\ref{eqn:split-path-integral}), 
with the boundary conditions $\phi(\pm R_\kappa)\,{=}\,f^{\Delta_-}\phi_{\kappa}^\pm$ on the two components of the 
boundary and $f\,{=}\,l/L\sech(R/L)$.
On the right hand side $\mathcal O_\pm$ denotes the corresponding dual operators of the CFTs at $\pm R_\kappa$, where
the rescaled bulk metric $\overline g:=f^2g$ naturally induces AdS$_d$ metrics.
The starting point to obtain concrete dS/dS correspondences is Vasiliev's minimal higher-spin gravity
on that AdS$_{d+1}$ bulk geometry with $d\,{=}\,3$, which is then dual to the \O{N} vector model on 
the two copies of AdS$_d$ constituting the conformal boundary.
With $\kappa\,{\leq}\,L$ the cut-off version of that duality (\ref{eqn:AdSAdS/cutoffCFT}), which corresponds 
to introducing a symmetric UV cut-off in both CFTs, provides a holographic relation defined on a part of the green 
shaded region in Fig.~\ref{fig:adsads}.
We can thus apply the analytic continuation (\ref{eqn:wick-dsds-adsads}), which realizes the transformation from AdS to dS used 
in \cite{Anninos:2011ui} just adapted to our choice of coordinates, resulting in
\begin{align}\label{eqn:dSdS/cutoffCFT} 
\Psi_\text{IR}^\mathrm{dS/dS}[\phi^{+}_{\kappa},\phi^{-}_{\kappa}]:=
 \Psi_\text{IR}[\phi^{+}_{\kappa},\phi^{-}_{\kappa}]\Big\vert_{(\text{\ref{eqn:wick-dsds-adsads}})}
 =\left\langle \exp\left\lbrace \ii \int_{\mathrm{dS}_d} \!\big(\phi^{+}_{\kappa}\,\mathcal O^{}_++\phi^{-}_{\kappa}\,\mathcal O^{}_-\big)\right\rbrace\right\rangle_{\text{CFT},\Lambda_\kappa,N\rightarrow -N}~.
\end{align}
The bulk geometry is transformed to the dS$_d$ slicing of dS$_{d+1}$ and the CFT metrics accordingly to dS$_d$.
We have thus obtained a duality of higher-spin theory on the dS/dS bulk geometry to a pair of dual cut-off CFTs on dS$_{d}$,
again with the continuation from \O{N} to \Sp{N}.
The dS/dS duality (\ref{eqn:dSdS/cutoffCFT}) is similar to the static-patch duality (\ref{eqn:dS/cutoffCFT}):
both involve higher-spin theory in the bulk and cut-off versions of the \Sp{N} CFT$_3$ on the boundary.
The different bulk geometries are reflected in the fact that the boundary theories are defined on
different spaces as well -- on a cylinder in (\ref{eqn:dS/cutoffCFT}) as compared to two copies of dS$_d$ in (\ref{eqn:dSdS/cutoffCFT}) --
and with different cut-off implementations.

In the AdS/CFT duality (\ref{eqn:AdSAdS/cutoffCFT}) we had intentionally left the choice of boundary conditions
for the CFTs on AdS$_d$ unspecified.
For each choice of CFT boundary conditions or $\mathbb{Z}_2$ identification on the AdS side the analytic continuation yields
a corresponding dS/dS duality (\ref{eqn:dSdS/cutoffCFT}).
The limit where the entire bulk geometry is recovered calls for a special treatment, analogously to
the discussion for the static patch in Sec.~\ref{sec:holography-for-static-patch}.
In that context the boundary conditions (\ref{eqn:static-patch-bc}) appearing naturally at the horizon 
may be of interest. We expect something similar to appear for dS/dS, in particular also
for the bulk metric. As the horizon boundary conditions are not pure Dirichlet, the 
boundary cut-off CFTs would naturally be coupled to dynamical gravity, as anticipated in \cite{Alishahiha:2004md,Alishahiha:2005dj}.
We leave more detailed investigations for the future and note that the picture
nicely agrees with the general discussions of dS/dS so far. 
It substantiates the discussion by providing a concrete example
and also a new perspective on the expected features.

We have seen that one and the same field theory, formulated on different background geometries
and with different cut-offs, can be dual to different regions of dS. 
Detailed discussions were given for the static-patch holography and for elliptic dS/dS,
and related dualities for other patches of dS can be 
obtained by similar analytic continuation.
It would be particularly interesting to study in more detail the static dS/dS slicing,
where the pair of dual cut-off CFTs in the corresponding AdS/CFT picture is defined with a 
spatial cut-off on AdS$_d$ in addition to the UV cut-off.
Since the bulk geometry is static in that case, this setting allows for a direct comparison to the
static-patch holography of Sec.~\ref{sec:staticpatch-holography}.

\section{Discussion \& Outlook}\label{sec:discussion}
We have argued that the dynamics of minimal higher-spin gravity on the dS$_4$ static patch, optionally  with a
radial cut-off, is 
encoded in a cut-off version of the \Sp{N} CFT$_3$ of anticommuting scalars on the Lorentzian-signature cylinder.
The discussion was based on a relation of the dS static patch to an inner shell of AdS via double Wick rotation
and a corresponding analytic continuation in the dynamics of the bulk and boundary theories.
As discussed in Sec.~\ref{sec:holography-for-static-patch} 
the limit where the dS bulk geometry becomes the entire static patch has to be taken carefully. 
The spatial part of the boundary cylinder straightforwardly arises as a holographic screen, while the time direction
only arises naturally in Fourier space.
The proposed duality allowed us to transfer lessons learnt from AdS/CFT to the description of (quantum) gravity on the static patch.
With the concrete dual description in terms of a cut-off CFT on the cylinder we have derived the number of 
degrees of freedom on the dS static patch from the dual theory. 
It is finite and respects a holographic bound, and the corresponding entropy reproduces the functional form of the 
thermodynamic horizon entropy.
To make the discussion more explicit we have then studied the two-point functions of the boundary theories on cut-off dS and AdS.
We found them related by the expected analytic continuation, which,
reflecting the different but related cut-off interpretations in the boundary theories, did not solely affect their normalization.
For the entire static patch as bulk geometry we have recovered the spectrum of quasinormal frequencies from the correlators of the
boundary theory on the horizon, and as a limiting case we have also recovered the proposal of \cite{Anninos:2011ui}.
Although the explicit discussion was limited to free bulk scalars we expect the established analytic continuation from 
AdS to dS to extend to perturbatively interacting fields.
We have then derived a similar relation of the geometries underlying the dS/dS correspondence to an inner
part of AdS, which allows to similarly provide an explicit realization.
It also results in a coherent picture of dS holography identifying the various incarnations as 
different forms of cut-off AdS/CFT.
 
There are also open questions which we think would be interesting to study in the future.
As discussed in Sec.~\ref{sec:scalar-explicitly-staticpatch} the boundary conditions arising naturally on the horizon 
of the static patch can be translated to AdS, and the two settings are then connected by analytic continuation.
However, it would be of interest to find an independent interpretation of these boundary conditions intrinsically
on AdS.
In that context alternative interpretations of the AdS Dirichlet problem may be relevant:
since a boundary condition at fixed $r_\kappa$ uniquely determines the bulk solution, one may
identify a localized source for the cut-off CFT at $r_\kappa$ with a non-local source for the 
dual operator in the full CFT, or vice versa.
A possibly related issue discussed briefly at the end of Sec.~\ref{sec:cut-off-AdS-CFT-Wick}
is the unitarity of the boundary theory and the question whether it is indeed restored by the
mandatory cut-off.
As an extension of the discussion here it would certainly be of interest to include fields of higher spin, in particular
a dynamical bulk metric. To this end it would be desirable to have a characterization of the spacetimes where the above 
construction can be carried out, as available e.g.\ for AdS/CFT in the form of asymptotically-AdS spaces.
It may also be possible to obtain further concrete examples by applying the discussion to other AdS/CFT dualities. 
We note in that context that on the group-theoretic level a similar analytic continuation as from \SO{N} to \Sp{N}
is also available for \SU{N} \cite{Dunne:1988ih}.
The crucial point will certainly be the extension of the analytic continuation from AdS to dS 
to the actual bulk dynamics, apart from which the discussion was pretty general already.
That may be possible for other variants of higher-spin theory, as relevant e.g.\ for the minimal-model holography \cite{Gaberdiel:2010pz}.

\begin{acknowledgments}
We thank Thorsten Ohl for useful discussions.
The work of AK is supported in part by the US Department of Energy under grant number DE-FG02-96ER40956.
CFU is supported by Deutsche Forschungsgemeinschaft through the 
Research Training Group GRK\,1147 \textit{Theoretical Astrophysics and Particle Physics}.
\end{acknowledgments}

\bibliography{dsds.bib}
\end{document}